# THE ROLE OF THE SATELLITE IN QUANTUM INFORMATION NETWORKS

Luca Paccard[1], Valentin Leloup[1], Luca Lazzarini[1], Agathe Blaise[2], Mailys Guerault[1], Mickael Faugeron[1], Fabrice Arnal[1], Mathieu Bertrand[1], Raphael Aymeric[1], Michel Sotom[1], Stéphanie Molin[2], Patrick Gélard[3], Pierre Besancenot[1], Cyrille Laborde[1], Laurent de Forges de Parny[1] and Mathias van den Bossche[1]

*[1]Thales Alenia Space, 26, Avenue J-F Champollion, 31037, Toulouse, France*
*[2]Thales SIX GTS, 4 Avenue des Louvresses, 92230 Gennevilliers, France*
*[3]Centre National d'Etudes Spatiales, 18 Av. Edouard Belin, 31400, Toulouse, France*

(contact: laurent.de-forges-de-parny@thalesaleniaspace.com)

**ABSTRACT**

Quantum Information Networks (QIN) attract increasing interest, as they will enable interconnection of multiple quantum devices in a distributed organization thus enhancing intrinsic computing, sensing, and security capabilities. The core mechanism of a QIN is quantum state swapping, based on teleportation, which consumes quantum entanglement, and which can be seen in this context as a new kind of network resource. The satellite is expected to play a central role for supporting global connectivity in such novel networks in which ground fiber links have stringent restrictions in length due to the absorption losses in optical fibers. There is indeed fundamental limits in the maximal fiber links distance which may not be exceeded for any unitary links. In this paper we clarify our motivations to develop such networks with satellites, and we discuss their associated use cases based on entanglement distribution, and we present the future potential users. We also assess quantitatively the ranges for which the satellite becomes mandatory in quantum information networks.

## 1. INTRODUCTION

The late 20th century was marked with the arrival of new technologies that manipulate and control quantum objects, now commonly referred to as the second quantum revolution. Manipulating quantum objects involves modifying their quantum states, which encode information at the fundamental physical level. Key properties such as state superposition, entanglement, the no-cloning theorem and quantum teleportation, play significant roles in these technologies. Quantum information networks (QINs) will enable the entanglement resource to be propagated (via entanglement swapping) over long distances, so that end-users can exchange quantum information via quantum teleportation. This field attracts increasing interest, as QINs would allow unprecedented computing, sensing, and security capacities and services.

The fundamental resource of QINs is the quantum entanglement. Quantum entanglement applies to a group of particles being generated, interacting, or sharing spatial proximity in such a way that the quantum state of each particle of the group cannot be described independently of the state of the others, including when the particles are separated by a large distance, as demonstrated with the use of the Micius satellite [1][2]. A QIN is a specific network concept which relies on generating, distributing and routing entanglement of bipartite quantum systems between many users.

The distance between QIN users has a significant impact on communication performance, especially when considering global-scale distances. Quantum signals are inherently weak and cannot be amplified due to the no-cloning theorem, imposing limitations. Fiber links have practical distance constraints, while free space links enable longer communication distances. In order to extend the connectivity to remote areas and allow global connectivity, satellite nodes are mandatory in such QIN, hence the involvement of Thales Alenia Space in the development of these novel systems.

This article highlights our motivation to develop QINs, particularly the space segment required for a global connectivity. We recall the main quantum use cases based on quantum entanglement involving distributed quantum computing and distributed quantum sensing, and which will benefit from a QIN to network them. The potential user, and their specific use cases, are discussed. These discussions are illustrated by high-level QIN architectures in order to emphasize the involvement of satellite(s). Finally, we present a rough assessment of the performance of ground and space links in a QIN to characterize the distance ranges for which the satellite becomes relevant – on the technical performance perspective.

## 2. QUANTUM ENTANGLEMENT-BASED COMMUNICATION USE CASES

In this section, we review the main quantum communication use cases based on the manipulation of entangled bipartite qubit system. This non-exhaustive list will give a brief idea of the potential of using entanglement between remote parties. The most promising use cases rely on networking of remote quantum machines such as quantum computers and quantum sensors. Therefore, we first start by a brief recall of the basics on quantum computing and quantum sensing.

*Quantum computing:*

Quantum computing is based on successive unitary operations, dictated by a quantum algorithm, that allow manipulating single or entangled qubits located in a physical qubit register over time. By exploiting state superposition and quantum entanglement manipulation, a quantum computer can realize, in principle, logical operations with a less significant effort (in time and number of operations) than a classical computer. The well-known examples are Shor and Grover algorithms, with key computing benefits, but also coming with threats with respect to cybersecurity. Different types of quantum computers exist, all facing the same main challenge: isolate the qubits to prevent them from the decoherence effect and maintain their quantum properties. Depending on the quantum computer, the qubit could have different nature, e.g. photonic, ionic, atomic or superconducting qubits. It is interesting to focus on a typical quantum computer architecture to understand the benefits and the technical limitations to overcome on quantum computing. Let us focus on what is a quantum computer, taking the example of a superconducting gate-based quantum computer[1]. In practice, a quantum computer is typically assisted by a classical computer and external control/readout electronics that allow for driving the quantum chipset (Figure 2-1 (a)). The quantum chipset (quantum processor) is typically a cooled electronic chip containing the physical qubits with microwaves connections for driving the logical quantum gates (X, Y, Z, S, T, R, Hadamard, CNOT, Toffoli, SWAP, etc.), for performing operations on qubits (Figure 2-1 (b)). The operations are typically rotating the qubit on the Bloch sphere or entangling the qubits together, see example in Figure 2-2.

---

[1] Other types of quantum computer exist, e.g. ionic or photonic quantum computers.

The quantum processor is located at the very bottom of the chandelier at temperature close to a few mK, in vacuum with thermal and external magnetic field isolation (Figure 2-1 (c)). The very cold temperature ensures maintaining the quantum coherence between the qubits. In this type of architecture, the main engineering challenge consists in increasing the number of logical qubits due to the required cryostat cable per qubits (e.g., liquid He4/He3, dilution refrigeration) for maintaining a very low temperature, and ensuring thermal dissipation for the sake of the qubit's stability and coherence. Increasing the number of logical qubits would also increase the complexity of the control electronics required for driving the logical gates.

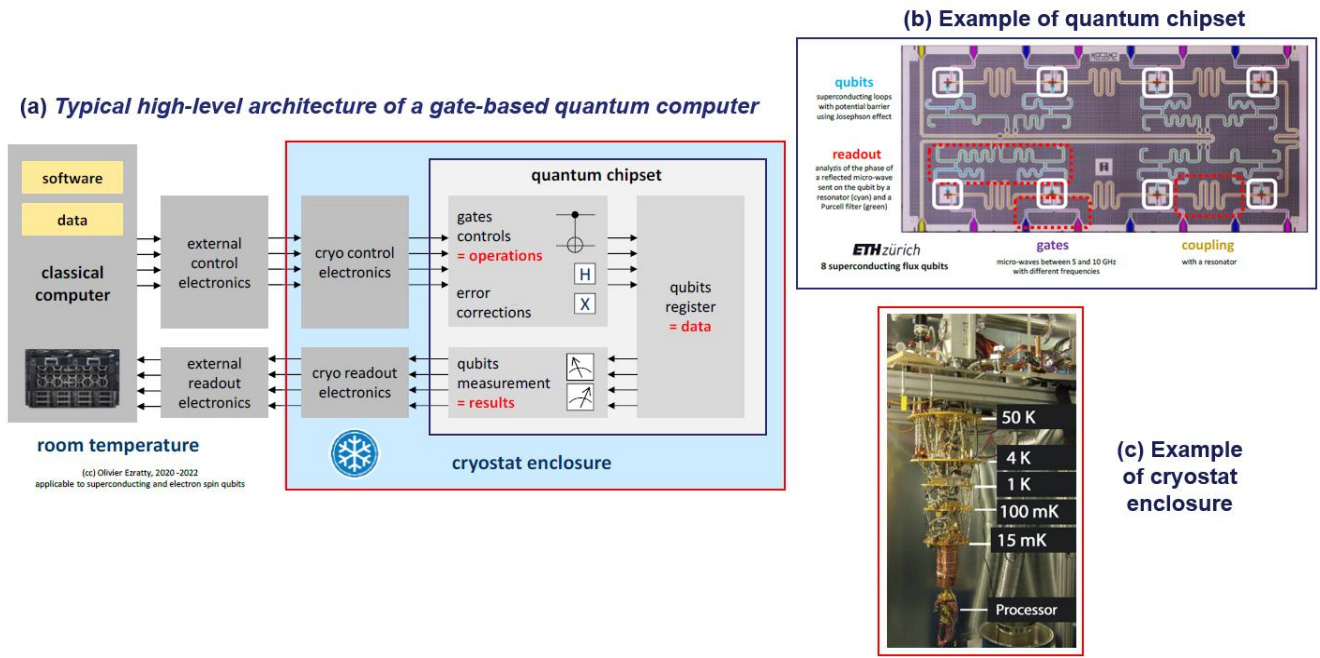

Figure 2-1: High-level quantum computer architecture, from [3]. (a) Typical high-level architecture of a gate-based quantum computer. (b) Example of a quantum chipset of 8-qubit superconducting processor from ETH Zurich showing its various components controlling the qubits. Source: The European Quantum Technologies Roadmap, 2017 and the thesis Digital quantum computation with superconducting qubits by Johannes Heinsoo, ETH Zurich, 2019. (c) Example of a cryostat enclosure in a chandelier architecture.

It is important to notice that quantum computing will probably not replace most of the use cases of classical computers. So far, quantum computing use cases are strongly architecture dependent and a universal[2] quantum computer does not exist today. However, quantum computing can bring a significant value for specific cases, like complex combinatorial problems, optimization problems, quantum physics simulation (e.g., solving molecular Hamiltonian), machine learning problems and fast integer factoring, e.g. prime number factorization in the Shor algorithm.

---

[2] i.e. capable of computing any computable sequence

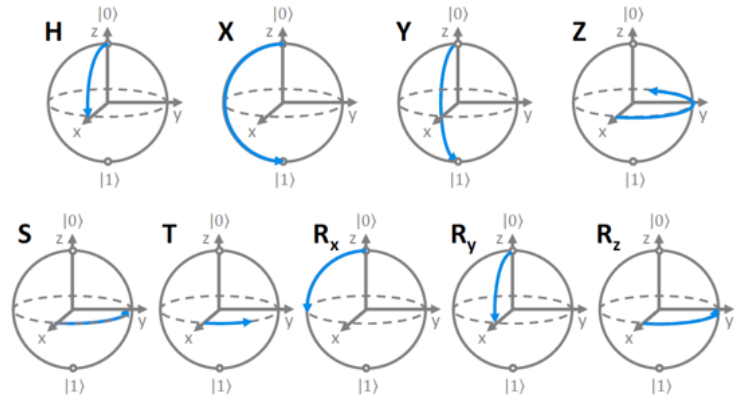
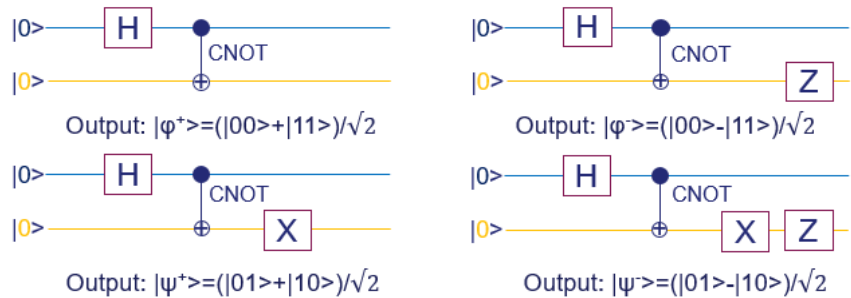

Figure 2-2: Bloch sphere representation of H, X, Y, Z, S, T and R single-qubit gates, from [3]. The gates H, X, Z and CNOT can also be used for the generation of the (maximally entangled) Bell states.

The quantum computing resource is characterized by a quantity called quantum volume, which takes into account both the Hilbert space of the useful qubits and the part of the resource devoted to real time error correction (flip or phase error) during a calculation.

When we consider two (classical) binary computers, the total computational resource of the system made up of the computers put together in a network is equal to the sum of the computational resources of these computers taken individually. Reversely, when we consider two quantum computers, the quantum volume of the system constituted by the two networked quantum computers is equal to the product of the quantum volumes of these computers taken each one in isolation[3]. It will thus be quickly interesting to put quantum chipsets in network, because the gain in performance will be exponential. Furthermore, as mentioned above, increasing the qubit density on a chip in the cryostat enclosure (called scalability) is the main engineering bottleneck. These scaling constraints strengthen the idea to physically fragment a large single quantum chipset into an area of smaller quantum chipsets, remotely connected by an entanglement distribution network, see discussion on distributed quantum computing and the use of the satellite.

*Quantum sensing:*

Quantum sensors are devices that leverage the quantum properties of a system to measure forces, fields, or time. The most striking example is the atomic clocks that allowed revolutionary advances in communications and absolute positioning systems such as GPS, among others [4]. The basic value of quantum sensors can be divided into two categories: those that may provide novel capabilities and/or performance levels not available with classical state of the art sensors (e.g. atomic clock); and those that may provide comparable capabilities and performance to existing sensors but in a more compelling size, weight, power, or cost envelope.

---

[3] Assuming a full connectivity of all qubits.

The Quantum Economic Development Consortium (QED-C)[4], that aims to enable and grow the quantum industry, has classified the quantum sensor types as a function of their category, current maturity and relevant applications (Table 1, from [4]).

**Table 1: Quantum sensor classes, their current state of development, and relevant applications**

| Category | Sensor type | Current maturity | Relevant applications |
|---|---|---|---|
| Timing | Microwave atomic clocks | Commercially available; broadly deployed | GPS-denied timing; Secure & resilient communications; Advanced sensing concepts (e.g., radar, reflectometry) requiring synchronized distributed sensor nodes |
| Timing | Optical atomic clocks | Research grade, including both laboratory-scale and portable designs | GPS-denied timing; Secure & resilient communications; Advanced sensing concepts (e.g., radar, reflectometry) requiring synchronized distributed sensor nodes |
| Inertial | Atom interferometer | Advanced prototypes, with limited commercial availability for gravimeters | High-end tactical grade and above inertial navigation |
| Inertial | Nuclear magnetic resonance | Advanced prototypes | High-end tactical grade and above rotation sensing |
| Inertial | Solid-state defect | Research grade | Space-constrained rotation and orientation sensing |
| Magnetic field | Atomic vapor | Commercially available | Resource exploration; underground infrastructure mapping; GPS-denied navigation; biomagnetics |
| Magnetic field | Solid-state defect | Primarily research grade; limited commercial availability of scanning microscope & bulk magnetometry devices | Magnetic microscopy; Widefield magnetometry; Geo-survey; GPS-denied navigation |
| Magnetic field | SQUID | Commercially available | Biomagnetics; resource exploration |
| Electric field | Rydberg sensor | Primarily research grade; custom sensors available for purchase from limited vendor base | Antenna calibration and other near-field spectrum sensing with mutual interference mitigation; wideband communications and sensing |
| Electric field | Solid-state defect | Primarily research grade | Electric field nano-microscopy; RF spectrum analysis |

---

[4] [QED-C | The Quantum Economic Development Consortium (quantumconsortium.org)](https://quantumconsortium.org)

The conclusion of Table 1 is that quantum sensors could be of very different types, for a wide spectrum of applications and maturity. Some quantum sensors devices available in the Lab or in the Market are presented in Figure 2-3.

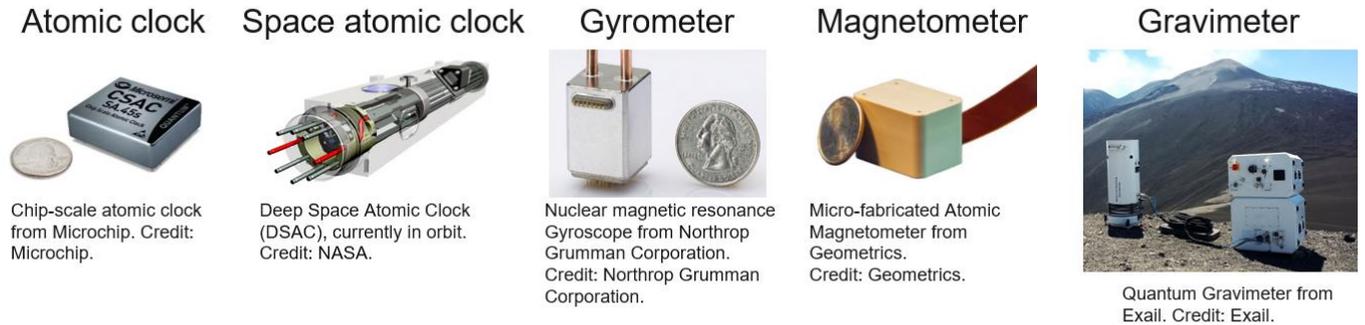

Figure 2-3: Example of quantum sensors devices available in the Lab or in the Market (figures taken from [4]).

Although the target applications of quantum sensing are very different from those of quantum computing, the idea and therefore, integration of networking quantum sensors is still interesting. The reason is similar: increase the performance due to the exponentiation of the quantum resource. In addition, the possible handling and processing of quantum sensor resources by quantum computers could also be imagined.

**Distributed quantum computing/sensing**:

Increasing the qubits density in a quantum computer chipset or in a quantum sensor, a principle called scalability, is a technological bottleneck actively studied. Scalability is mandatory to reach a satisfactory performance in quantum computing to run most of the known interesting quantum algorithm, e.g. Shor algorithm. At the moment, building a large-scale quantum chipset is still a very hard task. This bottleneck could be overcome, in principle, by interconnecting several quantum devices containing fewer qubits. Collectively using quantum computing or sensing devices that are connected in a network is known as distributed quantum computing or sensing for which quantum communication is essential [5]. This principle is depicted in Figure 2-4 where two quantum computers or sensors are connected via a quantum information network distributing quantum entanglement and, thus, allowing for the combination of their computational power.

In Figure 2-4 we assume a fully connected architecture for each individual quantum computer or sensors (all qubits of the same chipset can interact with any other, similarly to Ref [5]). The resulting distributed entangled resource is partially connected through qubits 2 and 8 only, which need to mediate the interactions between other remote qubit pairs, driven by a quantum algorithm with local and remote quantum gates [6]. The photonic quantum network allows for this remote connection by distributing entangled photon pairs and by using transducers (orange area in Figure 2-4) at the quantum network interface to enable interactions between different physical qubit encodings (e.g. for photonic to superconducting qubit conversion). In our proposed vision, the quantum information network comprises ground and space photonic entangled photon sources that generate the entanglement inside sections of the network, and Bell state measurement nodes that ensure the swapping of entanglement over the different sections, necessary for the establishment of an end-to-end entanglement link. We also assume the use of quantum memories for both storage and time synchronization purposes. The use of a space photon entangled source on-board a satellite allows for an extension of the scale of the quantum information network (see Section 4).

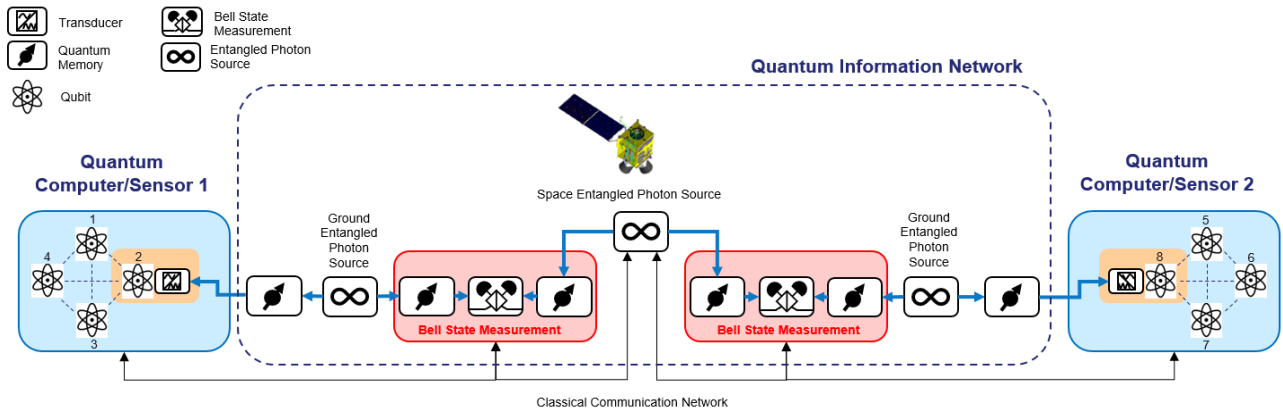

Figure 2-4: Scheme of two quantum computers or sensors interconnected via a quantum information network, allowing for the combination of their computational power.

The quantum information network can be replaced by a simple quantum link (e.g. an optical fiber, or a single entangled photon source) if the quantum registers are close enough. However, as explained in Ref [6], the interconnection of registers via entanglement is preferable rather than direct state transfer due to excitation loss (scattering/absorption) associated with the optical channel. The use of Bell state measurement for entanglement swapping is intrinsically more robust against excitation loss in the channel, since it only reduces the success probability. As explained in Ref [6], entanglement can then be used as a resource to teleport a quantum state from one sub-system to another. More generally, entanglement provides a physical resource to implement non-local unitary coupling gates, such as CNOT gate, see Ref [6] and reference therein.

**Blind or delegated quantum computing**:

The interest of this application is to make quantum computing accessible to small structures that cannot afford a large computer, while guaranteeing the confidentiality of the results obtained. To do so, an initial state is prepared in a "light" client quantum terminal, along with the configuration of quantum logic gates (i.e., inputs of the quantum-computing program and the quantum-computing program itself). These states and gate sequence are transmitted to a quantum server in a computing center that executes the sequence of logic gates on the initial state before returning the resulting final state to the client terminal (Figure 2-5) [7]. This computation is blind because it does not work if an actor tries to measure the information transmitted between the server and the quantum terminal: a measurement would disrupt the state and make both the computation false and the spy visible. In particular, even if it knows the sequence of logic gates, the computing center does not know the inputs/outputs of the calculation. This is discussed in Ref. [8] under strong assumptions concerning the entanglement of the servers.

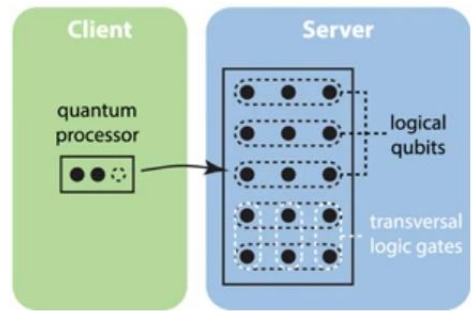

Figure 2-5: An illustration of blind quantum computing in which the client has access to a quantum computer capable of performing arbitrary operations on a constant number of qubits, from [7].

**Large base quantum interferometry**:

It consists in gathering quantum signals collected in distant locations, and routing them to a common location without measuring them, combining these gathered signals, and then measuring the state of the combination, in order to extract extremely fine differential information, or information about the quantum correlation of the signals [9]. This principle has been experimented in the context of astronomical radiation sources [10]. Other applications are also proposed for combining the output states of cold atom interferometers measuring for example the gravitational field (chrono-gravimeters) [11]. Such uses could eventually go beyond the field of fundamental or operational science. In remote sensing, the use of coherent sources could give rise to quantum detection of coherent photons emitted by the same scene at several locations, and allow a new type of images. One can also imagine quantum LIDAR arrays that exploit coherence. The quantum state of each collected element must be transferred via a quantum information network from the sensors to a common location to be jointly measured.

**Quantum time transfer**:

It consists in preparing distant quantum systems in states whose temporal evolution is known, but not the offset of one with respect to the other (e.g. clocks). A protocol, implemented thanks to the interconnection of the two distant quantum systems, and allowing the entanglement of the states of the two distant clocks allows to synchronize the evolution of their respective states. The observation of one of the clocks allows it to be synchronized with respect to the other, a priori without any other limit of precision than the control of the measurement, and this despite the delocalization [12]-[16].

**Quantum Key Distribution**:

This category of use cases involves securing communications. First of all, the implementation of advanced quantum key establishment protocols also known as Quantum Key Distribution (QKD), then means of distribution of encryption material, verification or authentication of data, and finally means of detecting failures or malicious actors in the network. Entanglement-based QKD key establishment consists of establishing shared confidential random sequences at two nodes in a network by exploiting quantum entanglement correlation. The Ekert91 [17], BBM92 [18] and MDI [19] protocols exploit these correlations in their own way. Once these sequences are established, they can be used as encryption keys by symmetric algorithms (disposable code, AES, Vernam, etc.). Ref. [20] provides an example of an operational entanglement based QKD on the ground over 50 km of real-field optical fibres and Refs. [1] and [2] are the state-of-the-art for an application involving a satellite (Micius). The keys established by the afford mentioned quantum protocols are a useful complement to other (e.g. post-quantum cryptography[5]) means of protecting communications once quantum computers have made current asymmetric encryption methods obsolete. Indeed, the hybridization of physical keys and mathematical algorithms will allow having properties of defense in depth, persistent confidentiality, a security guaranteed by the laws of physics and *in fine*, to free oneself from conjectures on the robustness of post-quantum algorithms to advanced quantum algorithms.

These advanced quantum key distribution protocols (Ekert91, BBM92 and MDI) use quantum entanglement, unlike 'Prepare and Measure' (PM) protocols based on information encoding on single photon. PM protocols allow for establishing QKD keys at both ends of an elementary optical channel. If one wants to establish PM QKD keys between two nodes that are far away (e.g. > 150 km), one must split the channel by introducing intermediate trusted nodes. These intermediate trusted nodes are therefore a weakness in the system's security. Advanced protocols that use entanglement distribution will only use intermediate nodes, composed of entangled photon source and Bell state measurement device, to weave the end-to-end entanglement links ensured by a quantum information network. For instance, Figure 2-6 shows two end-users that receive end-to-end entanglement through a quantum information network, and generate QKD keys by exploiting the quantum correlation after photon measurements. Therefore, contrarily to intermediate nodes of PM protocols, cryptographic keys do not transit through these nodes when using advanced protocols, which is a great advantage for the global system security. Once the end-to-end entangled link is established, one can proceed to the establishment of QKD keys by exploiting end-to-end entanglement. The QKD keys appear in plaintext at the two end terminal nodes only, which eliminates the security weakness of PM QKD keys. Quantum data authentication or verification allows remote information contents to be compared without exchanging this information in any way. The first protocols in this field are quantum digital signatures and quantum watermarks.

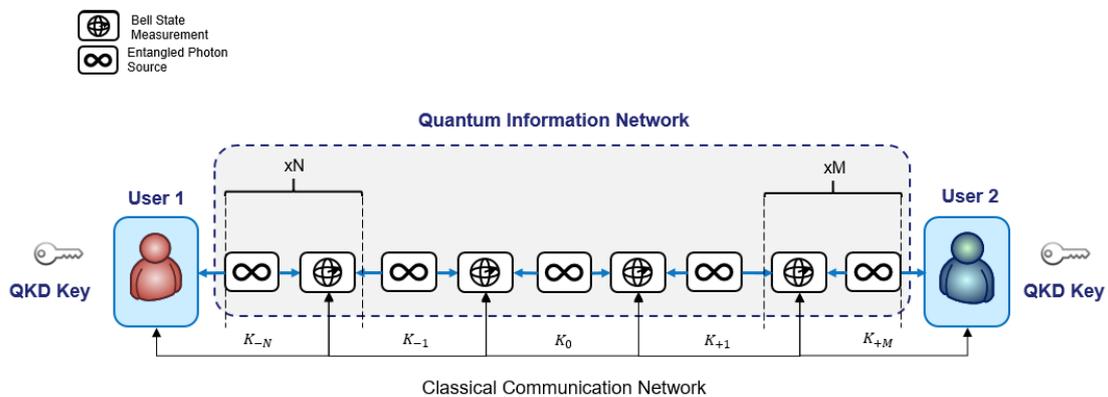

Figure 2-6: A quantum information network provides entanglement to two end-users. These users generate QKD keys by processing the measured correlations. The quantum information network does not contain trusted nodes, contrary to prepare and measure based protocols. The entangled photon source and Bell state measurement devices can be located on the ground or on board a satellite.

Secure communications can also benefit from the ability of quantum networks to self-detect failures and/or malicious actors within these networks. Fault detection will be critical for distributed computing application presented above. The ability to detect malicious actors will always be a benefit to the companies that exploit these QINs. The protocols that enable this use are known as *quantum byzantine* agreement protocols. They exploit entanglement to detect inconsistencies between states of parts of the network. This inconsistency being the signature of an abnormal behavior, its detection allows intervening quickly to isolate the defective or malicious element of the network.

---

[5] Post-Quantum Cryptography | CSRC (nist.gov)

## 3. POTENTIAL USERS

In this section, we try to give a hint of the potential users per sector that could benefit from the use of quantum information networks, based on Ref [21]. The activity sectors that could benefit from the services of a QIN are presented in Figure 3-1.

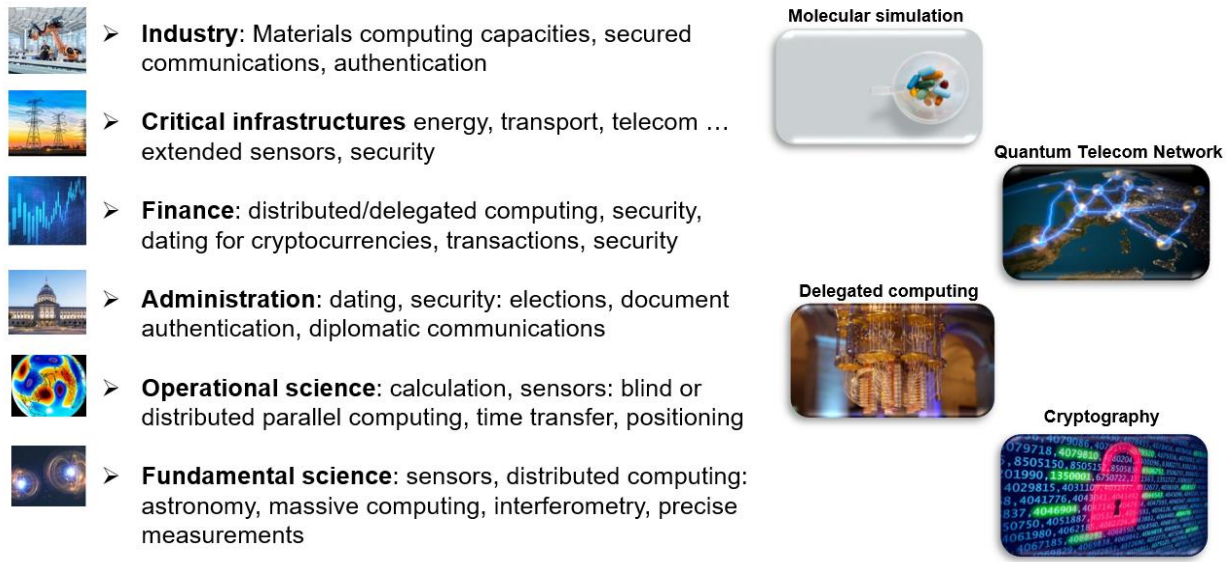

Figure 3-1: Example of activity sectors that could benefit from the services of a QIN [21].

In the following paragraphs, we explain how each sector could benefit from QIN services.

In the industrial field, a number of players have already expressed their interest in using quantum delegated or blind computing capabilities [22]. Their objective is generally to take advantage of the capabilities of these computers to solve optimization problems relative to their business sector, or to represent physical systems of interest more accurately. In this respect we mention: the mechanical industries, for example the automotive/aeronautical sector, for the optimization of finite element models; the pharmaceutical industry, to simulate larger and more complex molecules; the materials industry, in order to predict the properties of materials, which are ultimately related to quantum physics. In addition to these industrial sectors, the high-performance computing (HPC) capacity providers themselves, who will be able to offer quantum HPC services to industry to meet all of the above mentioned needs through adapted interfaces. Beyond blind computing, HPC actors will be interested in the capability to distribute computing tasks over several machines in order to leverage more processing resources. In addition to quantum computing, industries will certainly be interested in the possibility of establishing encryption keys to protect trade secrets.

Under the term critical infrastructure, we group together the various large systems deployed on the scale of countries or continents that provide the structural means for the functioning of a society, e.g., power generation and distribution systems (electric, gas, etc.), water distribution networks, public transportation systems, health data centres, telecommunication networks, and specific systems for managing critical infrastructures or monitoring the environment to anticipate major crises (such as flood control in The Netherlands, seismograph networks in Japan, pollution sensors, etc.).

Within these critical infrastructures, we have identified two types of use. On the one hand, many critical infrastructures need precise timing systems for their operations, and the required accuracy increases with the complexity of these infrastructures. This is either to coordinate operations at a very high frequency, or to be able to date events in order to identify incipient cascades of events that could lead to a system blockage. For instance, the growth in throughput of communication networks requires ever more precise time control of the sequence of execution of tasks at their nodes. Another example is the timing of abnormal voltage variations in an electrical distribution network, which makes it possible to cut off certain links if they can be identified, before the anomaly propagates until the network collapses. The increasing need for accuracy will eventually lead to accuracies that only quantum sensors will be able to satisfy, and quantum sensor networks will emerge naturally. On the other hand, these infrastructures all have a command/control service for their distributed elements, and the intervention of a malicious actor in these services can block the infrastructure, which is a way to jeopardise the society that benefits from this infrastructure. Protecting this service relies, among others, on securing communications (at least authentication), which a QIN will also be able to do. Major telecom operators seriously consider securing their networks in this manner.

Banks and the financial sector are critical infrastructures of a particular kind, in the sense that if they share the need for precise timing of transactions and the need for securing their communications, their general evolution leads them to use ever more computing resources. The key driver for this evolution is the movement of tokenisation of finance assets, whose most visible effect is the appearance, and then the generalization of cryptocurrencies—not the already existing private cryptocurrencies, but the ones that central banks are preparing, and which are planned to become the dominant means of exchange in our economies.

As part of such an infrastructure, QINs have two roles to play. On the one hand, enable distributed computing, as the authentication of transactions within a blockchain will most certainly be accelerated by the use of Grover-type quantum algorithms. One the other hand, enable the global security of processing and storage assets. Indeed, in such systems, the risk of a security breach that allows a malicious actor to steal or destroy value is no longer evaluated at the order of hundreds of millions of euros, as is the case with current attacks on Bitcoin. The level of risk of a security breach on central cryptocurrencies is on the scale of a country's economy or that of the entire planet. Such stakes are likely to attract players on a whole other level, with far greater means of attack than current cryptocurrency offenders. The direction that central banks are taking is to use only information-theoretically secured means, among them QKD and post-quantum keys.

Concerning administration, we group under this term the means at the disposal of the political management organizations of our societies. These administrations are in charge of very vast domains (elections, diplomacy, justice, army, etc.) that are likely to benefit from quantum telecommunication services. We can cite as examples, electoral processes, which could benefit from perfectly anonymous and perfectly reliable voting systems based on quantum protocols [23][24]—at least in democratic societies—, and authentication processes of digital documents for the validity of contracts, mandates and diplomatic communications (e.g., comparison of documents without exchange, encryption). These are example of applications of QINs that fall within the broader domain of secure and/or anonymous communications and that will go well beyond what is currently possible.

For operational science, we distinguish the case of operational science from that of fundamental research. While the latter is concerned with the fundamentals of our scientific knowledge, the former has as goal the implementation of systems that rely on advanced knowledge to provide information that can be used in the very short term by societies (e.g., meteorology, geodesy, ...). For operational science, one can envision a large number of use cases for a QIN: services that rely on heavy computational activity (resolution of optimization problems in geodesy, positioning or massive parallel computation in weather forecast) will certainly be interested in the use of distributed and probably also blind computation; entities providing time reference services will certainly use QINs to improve performance in terms of stability and accuracy of their services though a more precise synchronization of reference clock time, as well as of distributed time. Geodetic services will also be able to use clocks synchronized by such means to implement chronometric geodesy means; large baseline interferometric systems of electromagnetic or gravity quantum sensors to further increase their sensitivity for the benefit of remote sensing and environmental monitoring.

Finally, the field of fundamental science will probably make use of all possibilities offered by QINs, since this activity is so demanding in terms of processing resources, measurement and detection, and means of communication. On the other hand, it is difficult to be more precise in the case of such an evolving, innovation-rich, and diverse domain, in particular on the time horizon that concerns the operational implementation of a QIN. It should be noted that the QIN itself is likely to become a means of scientific observation, like a large distributed instrument, because of its great sensitivity to very subtle phenomena [25].

## 4. GLOBAL QUANTUM CONNECTIVITY WITH SATELLITES

In this section, we give a rough assessment of the service level, quantified by both the number of entangled photon pairs received by two end-users and the associated fidelity, provided by two segments: the ground and space segment. We consider a scenario with two paths, the ground path and the space path, and assume both paths are available simultaneously, meaning the satellite has visibility of the two ground stations at the same time (Figure 4-1). The end-users A and B receive simultaneously entangled photon pairs from these two paths, without routing nor scheduling consideration between the two paths.

The questions we address here are:
- How many entangled photon pairs can be sent and received by the ground and space path over time, depending on the distance between users A and B?
- What is the distance for which the satellite becomes mandatory?

**Hypotheses**:

For this rough assessment, we assume:
- A satellite with an entangled photon source on-board, with double optical terminal and with suitable pointing, acquisition and tracking system. The satellite communicates with two optical ground stations, from time $t_i$ to time $t_f$, each station is equipped with a quantum memory.
- A ground segment composed by a chain of entangled photon sources, fiber links, and Bell state measurement devices. Each Bell state measurement device is equipped with two quantum memories, for time synchronization purposes. We assume a symmetrical configuration at the ground (source in between, balanced channel losses) in order to optimize the service performances.
- All entangled photon sources are identical
- All quantum memories are identical, with very optimistic parameters with respect to the state-of-the-art

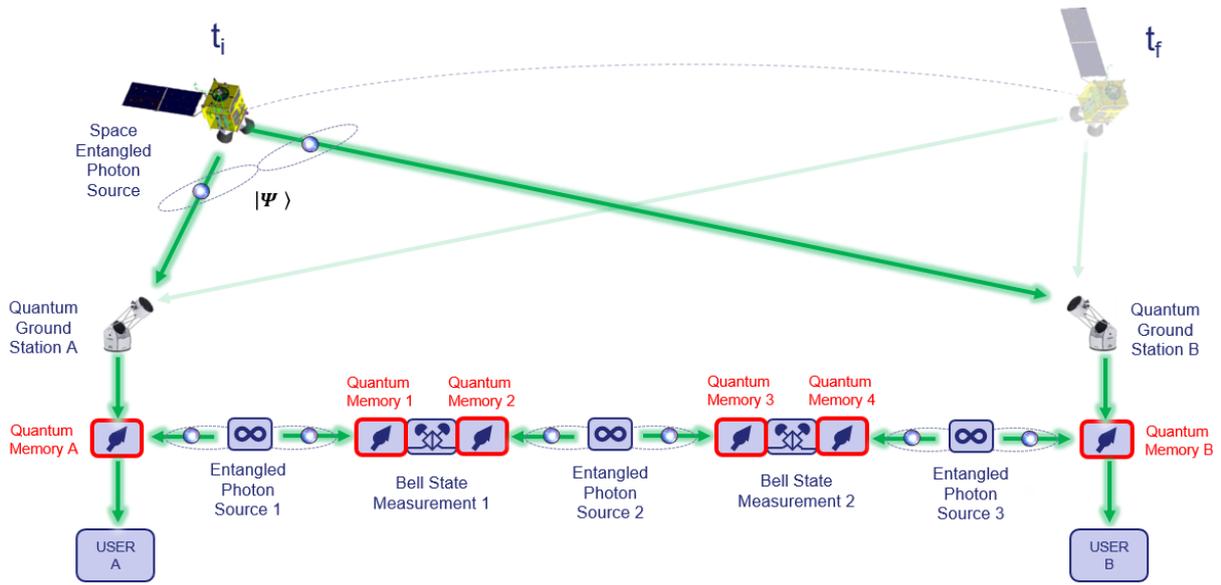

Figure 4-1: From time $t_i$ to time $t_f$, the users A and B receive quantum entanglement from both the satellite and the ground path. The level of service received by these two possible paths depends on the satellite's orbit (altitude, elevation, slant ranges, communication duration, etc.) for the space path and depends on distance between user A and B (number of relay nodes, quantum memories and entangled photon sources performance, optical fiber distance, etc.) for the ground path.

**Ground station locations**:

We consider three configurations for the location of user A and B, see Table 2.

**Table 2: Three configuration considered for location of user A and B**

| Configuration | User A | User B | Satellite | Distance A-B | Map |
|---|---|---|---|---|---|
| Short distance | Calern (FR) 43° 45' 13.2" N, 06° 55' 22.7" E | Haute Provence Observatory (FR) 43° 55' 51″ N, 5° 42' 48″ E | Inclination: 90° RAAN: 134.4° Altitude: 600 km | ~ 90 km | |
| Medium distance | Calern (FR) 43° 45' 13.2" N, 06° 55' 22.7" E | Toulouse (FR) 43° 36' 15.3" N, 1° 26' 37.2" E | Inclination: 90° RAAN: 132.5° Altitude: 600 km | ~ 440 km | |
| Long distance | Bordeaux (FR) 44° 50' 16.04" N, 0° 34' 45.048" W | Padova (IT) 45° 24' 28.69" N, 11° 53' 9.10" E | Inclination: 90° RAAN: 134° Altitude: 600 km | ~ 980 km | |

We have selected these station locations because most of them are equipped with existing optical ground station and this set allows for three range of distances: short (Calern-Haute Provence Observatory), medium (Calern-Toulouse) and large (Bordeaux-Padova). We expect a larger amount of level of service with the ground segment for short distance. Indeed, the performances of a Bell state measurement devices chain decreases with the number of those devices, thus with the distance. This decrease of performance is due to the quantum memories performances – although assumed to be above the state-of-the-art hereafter – and to the fact that a Bell state measurement device works successfully half of the time (at least with linear optics): only two Bell states over four can be used for the entanglement swapping. Moreover, for simplicity, we chose the Right Ascension of the Ascending Node (RAAN) of the satellite such that the distances between the satellite and each station are almost identical during the pass.

**Models for the ground and space path performance assessment**:

The models used for our simulations are detailed in the Annex paragraph for the sake of the clarity. The particularity of our models is that we consider multimodal quantum memories for optimization of the end-to-end entanglement swapping for the ground path. We also assume an optimized configuration for the ground path, e.g. balanced losses in the channel of each entangled photon sources (source in the middle).

Our models allow for the calculation of two key performance indicators we will compare for the ground and space paths:

- The number of received entangled photons at the end user-nodes (see Annex: $N_{pair,\ ground}$ Eq.(7) and $N_{pair,\ space}$ Eq.(16));
- The fidelity of the received entangled photons at the end user-nodes (see Annex: $F_{chain}$ Eq.(15) and $\mathcal{W}_{stray}$ Eq. (21)).

The list of parameters used in our simulations is also provided in Annex, Table 4. The parameters used for the quantum memories are significantly optimistic but seems a good target for an operational QIN. The orbital simulations are performed with ANSYS Systems Tool Kit®, a powerful commercial software enabling a mission scenario description including a detailed and realistic orbital propagation, taking into account, e.g., the non-sphericity of Earth, the residual atmosphere drag on the satellite as a function of its weight, geometry, and orientation (attitude sequences), as well as the ground station locations.

**Results**:

We start by focusing on the throughput, particularly by discussing the number of entangled photon pairs received at user nodes during the satellite pass. Figure 4-2 shows the instantaneous rate of entangled photon pairs received at user nodes and the integrated rate over time during the satellite pass. The longer the distance between users, the shorter the flyby time as the photons distribution requires simultaneous visibility of both ground stations by the satellite. The rate, and by extension the number of distributed pairs during the pass is also lowered as the link budget is affected by a higher slant range and lower elevation angle. The maximum rate of entangled photon pairs received at user nodes reaches ~12,100 (~2,500) for the short (long) scenario.

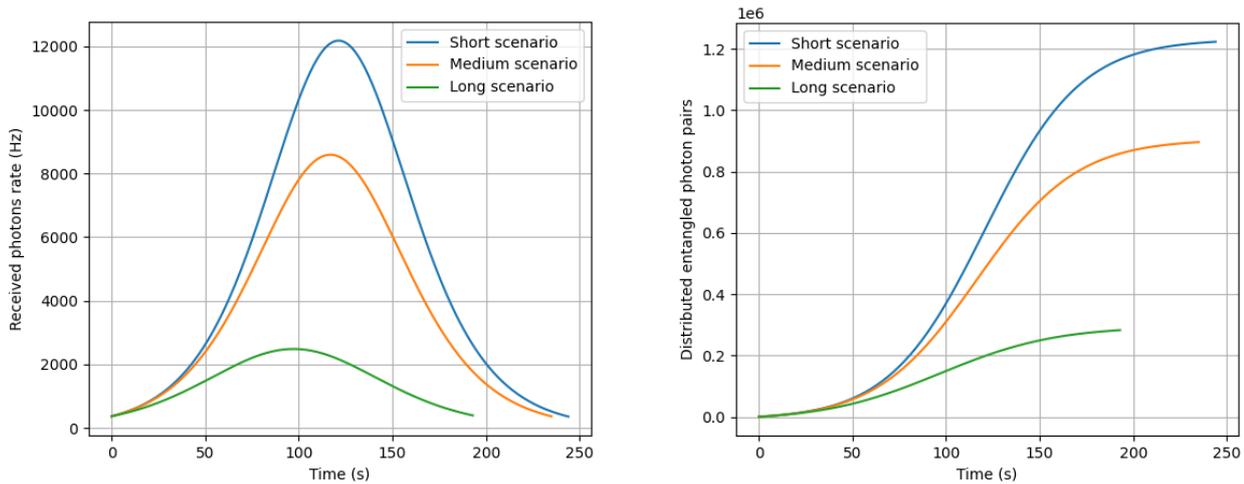

Figure 4-2: (Left) Rate of entangled photon pairs received at user nodes, distributed over time during the satellite pass. (Right) Integrated number of entangled photon pairs received at user nodes, distributed over time during the satellite pass.

Still focusing on the throughput, we now compare the space path performance with the ground path, considering only the integrated rate. In Figure 4-3 we compare the performances of a satellite pass with the distribution of entanglement through the terrestrial segment using three distinct methods:

- **Satellite**: Entanglement is distributed from the satellite to the ground stations (Figure 4-2 (right)).
- **Ground entanglement distribution without quantum repeaters**: This methods consists in using only a single entangled photon source located at the ground in-between the users and sends one photon to user A and one photon to user B, with fiber optic connections.
- **Ground entanglement distribution with optimized number of quantum repeaters**: This methods uses a number of quantum repeaters between end-users optimizing the rate of distributed entangled photons and respecting a criteria of fidelity, calculated with Eq. (15) (see Annex). The fidelity criteria must be considered because an optimization only relying on distribution rate tends to add a greater amount of quantum repeaters, which will have an impact on the end-to-end fidelity as the system will rely on more imperfect elements, each impacting fidelity.
  For a fidelity criteria of $F > 0.8$, the number of quantum repeaters optimizing the entangled photon pairs rate is:
    o Short scenario : 1 quantum repeater
    o Medium scenario : 6 quantum repeaters
    o Long scenario : 6 quantum repeaters

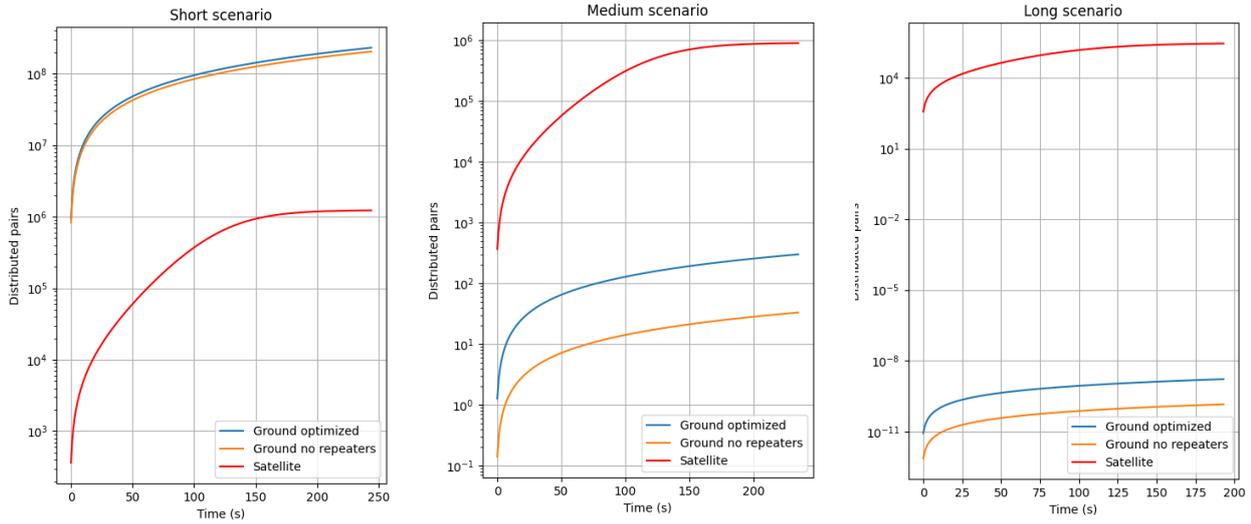

Figure 4-3: Comparison of the number of distributed photon pairs on the three scenarios (short, medium and large distances) with a fidelity requirement of $F > 0.8$ for the optimization of the ground network.

Comparing the throughput for the three scenario, we see that:
- Short scenario (Calern-Haute Provence Observatory, distance ~ 90 km): The ground path has the better performance (~2 order of magnitude compared to the space path), as expected, and the satellite has the worse level of service. Using one swapping node (quantum repeater) has better performance than plugging just one entangled photon source.
- Medium scenario (Calern-Toulouse, distance ~ 440 km): The satellite provides a level of service better than ~3 order of magnitude compared to the ground path with six swapping nodes. Clearly, using only one entangled photon source without swapping node is sub-optimized.
- Large scenario (Bordeaux-Padova, distance ~ 980 km): As expected, the satellite provides a level of service better than ~4.5 order of magnitude compared to the ground path with six swapping nodes. Here also, using only one entangled photon source without swapping node is sub-optimized.

The conclusion is that the use of the space path becomes significantly useful, compared to the ground path, for connecting users 400 km away.

As explained above, the throughput is not the only key driver. The signal quality, quantifies by the fidelity, is a second key performance indicator that should be assess when comparing two path of a quantum network. Indeed, two users can receive a high number of entangled photon pair but with a low fidelity, thus making the signal unusable.

The averaged fidelity of the distributed entangled photons of each scenarios, calculated with Eq.(15) and Eq. (19) (see Annex), are detailed in Table 3. For the space distribution of entangled photons, the only dynamic parameter taken into account is stray photons mixing with qubits from the satellite. However, the field of view used in our model takes into account fiber coupling at the ground level, which results in a very small percentage of stray photons entering into the quantum memories. The signal-to-noise ratio (i.e. number of useful photons versus total photon measurements) between photonic qubits sent from the satellite and stray photons is very high. The only elements degrading fidelity are thus source and quantum memory imperfection, as well as the freespace channel imperfection (misalignment). Concerning the ground path, the fidelity degradation is due to the quantum repeater imperfections, as well as the impairments in the optical fibers. The longer the fiber links, the more dark counts have an effect on the overall fidelity.

Table 3: Fidelity of distributed entangled photons for all three scenarios

| Scenario | Fidelity of ground distribution | Fidelity of space distribution |
|---|---|---|
| Short | 0.94 | 0.97 |
| Medium | 0.91 | 0.97 |
| Long | 0.81 | 0.97 |

The conclusion is that the space path allows a better conservation of the fidelity than the ground path, the latter involving a succession of devices (e.g. BSM, quantum memories) degrading the end-to-end fidelity. Indeed, only 2% degradation are observed for the space path for all scenario, the space entangled source fidelity being 99% in our simulations.

## 5. CONCLUSIONS

This article discuss a quantum communication topic of growing interest, the Quantum Information Network (QIN) which is much more general than QKD network applications, and for which the satellite has a crucial role to play. Quantum information network are designed to connect quantum computers or sensors, and can, in principle increase exponentially the overall capacities of such devices by putting them in network (i.e. increase computing power, greater measurement accuracy …). The core mechanism of a QIN is entanglement swapping to propagate entanglement to the users' access point, who consume this entanglement resource to teleport the quantum states they wish to communicate. Quantum entanglement can be seen in this context as a new kind of network resource. QIN aims to distribute to end-users quantum information over long distances via quantum teleportation or entanglement swapping, trough ground and space paths, and enabling the realization of the quantum use case.

In this article we clarify Thales Alenia Space motivations to develop such networks. We first discuss their associated use cases and the future potential users. We briefly discuss the quantum computing and quantum sensing topics, and explain why QIN are interesting for the exponentiation of the overall capacities. Particularly, QIN can be a mitigation strategy in quantum computing that suffer from the scaling problem, i.e. increase the qubit density in a single quantum chipset. QIN can overcome this problem by networking many quantum chipset with a small number of qubits. For sensing, the main benefit of QIN is similar: increase the performance due to the exponentiation of the quantum resource. We also present other use cases such as distributed quantum computing/sensing, blind or delegated quantum computing, large base quantum interferometry, quantum time transfer, and entanglement based-quantum key distribution. Then, we try to give a hint of the potential users per sector that could benefit from the use of quantum information networks. The activity sectors that could benefit from the services of a QIN are at least the following ones: industry (automotive, chemical, etc.), critical infrastructures, finance, administration, operational science and fundamental science. Their objective is generally to take advantage of the capabilities of these computers to solve optimization problems relative to their business sector, or to represent physical systems of interest more accurately. In addition to quantum computing, industries will certainly be interested in the possibility of establishing encryption keys, without trusted nodes, to protect trade secrets.

Finally, we have developed a mathematical model that allow for the assessment of a ground and space path in a QIN, considering multimodal quantum memories. We have considered realistic scenario with short (~ 90 km), medium (~ 440 km) and large (~ 980 km) distances for the assessment of the ground and space path performance with numerical simulations, assuming the two paths available. We calculate two key performance indicators: the number and fidelity of received entangled photons at the end user-nodes. According to our model, assumptions and simulations, we show that the space path becomes significantly useful, compared to the ground path, for connecting users 400 km away. Also, the space path allows a better conservation of the fidelity than the ground path, the latter involving a succession of devices degrading the end-to-end fidelity. These results should be considered as preliminary results since we are improving our models and parameter estimation. We remain open for discussion for improving these simulations.

## ACKNOWLEDGEMENTS

This work was supported by the French Space Agency, the Centre National d'Etudes Spatiales (CNES), the European Space Agency (ESA), and Thales Alenia Space. It relies on numerous prior works supported by the European Commission, the French Defence Innovation Agency (AID) and the French National Research Agency (ANR).

## ANNEX

In this annex we describe our models used for the calculation of the ground and space path performances. A table of the parameters used for our simulations is also provided.

**Model for the ground path**:

The advantage of a chain of quantum repeaters lies in the presence of heralded quantum memories within the nodes. These quantum memories make it possible to reduce the impact of losses in propagation channels (optical fibers for ground networks and free space for satellite links). Indeed, in the context of a simple chain as illustrated in Figure 5-1, if a photon is lost between a source and a node at a given timeslot but at this same instant another pair is received from a another source, then the latter can be stored in a memory while a photon from another pair is received.

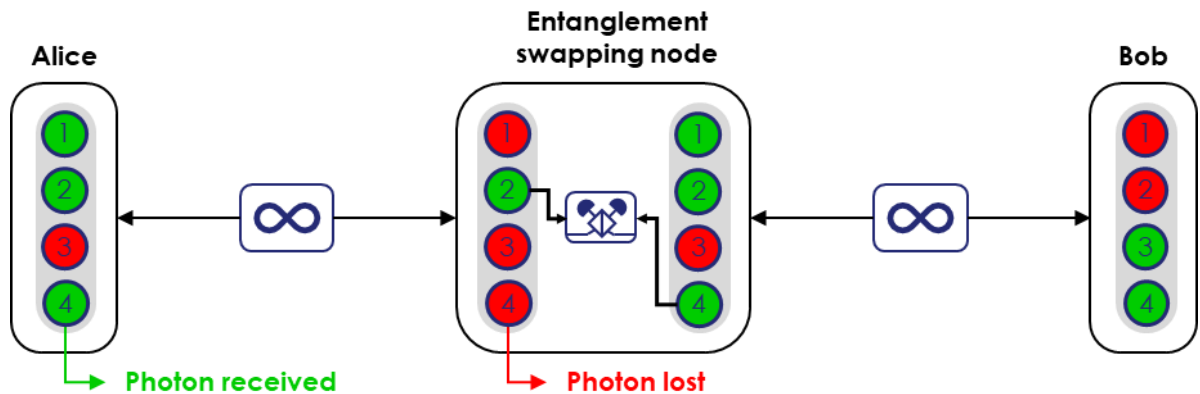

Figure 5-1: Bell state measurement device with quantum memories. Alice (Bob) share with the entanglement swapping node the pair #2 (#4).

Figure 5-1 shows nodes with multiplexed quantum memories. That is to say, they have several slots in which qubits arriving at different times can be stored. Alice and Bob have one memory with four slots and the entanglement swapping node has two memories with four slots each (or one memory with eight slots). The memories are indexed with timeslots, meaning that a photon may be written in the first memory slot during the first timeslot, etc.

A timeslot is defined as $\delta t = 1/R_{src}$, with $R_{src}$ the entangled photon pairs source pump rate and the duration of timeslot is defined as $t_k = k \cdot \delta t$ with $k \in [1, N]$ and $N$ the number of memory storage modes.

We see in Figure 5-1 that the entanglement distribution between Alice (resp. Bob) and the entanglement swapping node only worked for timeslot $k = 2$ (resp. $k = 4$). Entanglement swapping is thus possible between the qubits received at timeslots $k = 2$ (resp. $k = 4$), something that would not have been possible without quantum memories. This process is possible with the hypothesis that quantum memories have writing announcement (heralding) for each storage mode.

In the example of Figure 5-1, we studied the behavior of a small chain for four timeslots. To do this, it was necessary to take into account quantum memories with four storage slots. In order to generalize this example for $N$ temporal intervals we will consider quantum memories with $N$ storage slots. Each of these slot has its own performance « $\eta_{QM}(t_k)$ », i.e. a probability of successfully writing and then reading the quantum information placed there. Given that we study the system after temporal instants, the performance of the locations will therefore vary (Figure 5-2). During a timeslot, the memory efficiency is considered constant.

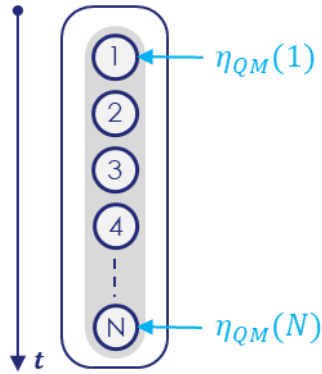

Figure 5-2: Performance of quantum memory locations

The probability of having correctly written and read the information stored in location #1 at the moment $N$ is lower than that of location #$N$. This is explained by the losses of coherence that are at play within a quantum memory: the longer information is stored, the lower the probability of recovering it. The efficiency of quantum memory locations is modeled to follow an exponentially decreasing trend over storage time of the type:

$$\eta_{QM}(t_k) = \eta_{QM,W} \cdot \exp\left(-\frac{t_k}{\tau_{QM}}\right) \tag{1}$$

where $\eta_{QM,W}$ is the writing efficiency of the memory and $\tau_{QM}$ is its characteristic storage time. We can then apply this formula for the different memory locations according to the desired temporal granularity. $\tau_{QM}$ is expressed as a number of timeslots as $\tau_{QM} = \tau_{QM,sec}/\delta t$ where $\tau_{QM,sec}$ is the characteristic storage time expressed in seconds and $\delta t$ the duration of one timeslot.

We assume that a chain of quantum repeaters can be divided into multiple elementary links. An elementary link corresponds to the distribution of entanglement from a source to two nodes, as illustrated in Figure 5-3.

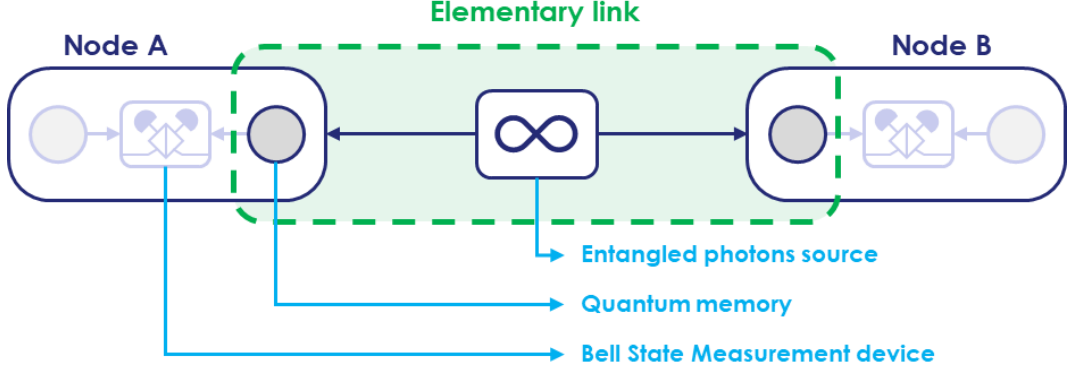

Figure 5-3: Elementary link of a chain of entanglement switches

The objective is to determine the probability of successfully transmitting at least one pair of entangled photons between the two nodes after $N$ timeslots. This objective corresponds to a worst-case scenario for system performances because if two pairs of entangled photons (or more) are received, only one will be kept to carry out the entanglement swapping. This scenario remains interesting for studying the optimal number of temporal instants after which to look at the state of quantum memories, and therefore optimize its use.

An elementary link is made up of three types of elements:
- A source of entangled photons;
- Two propagation channels (on either side of the source);
- Two quantum memories.

The entangled photon source is characterized by an entangled photon pair generation efficiency $\eta_{src}$ and a pump rate $R_{src}$.

The propagation channels are characterized by their length $l$ and their efficiency $\eta_{ch}$ which correspond to the efficiency of reception of a photon coming from the entangled photon source. This efficiency depends on the channel length for fiber optic propagation, and on both channel length and elevation angle for free-space propagation.

Quantum memories are characterized by the efficiency $\eta_{QM}(t_k)$ of their different modes, where $k \in [1, N]$, with $N$ the number of memory storage modes. $\eta_{QM}(t_k)$ is function of additional parameters such as the writing probability $\eta_{QM,W}$ and the characteristic time $\tau_{QM}$ (see Eq. (1)).

The probability of failing to transmit the entanglement between two nodes at timeslot $t_k$ is:

$$p_{fail}(t_k) = 1 - \eta_{src} \cdot \eta_{ch}^2 \cdot \eta_{QM}^2(t_k) \qquad (2)$$

The efficiency $\eta_{ch}$ is squared because it is equal on both sides of the source (under the assumption that the source is located halfway between the nodes for optimization purpose). The probability of failing to transmit the entanglement between two nodes after $N$ time instants is:

$$p_{all\ fail} = \prod_{t_k=1}^{N} \left(1 - \eta_{src} \cdot \eta_{ch}^2 \cdot \eta_{QM}^2(t_k)\right) \qquad (3)$$

Thus, the probability of transmitting at least one pair of entangled photons on an elementary link after $N$ timeslots is:

$$p_{success \geq 1(elem\ link),N} = 1 - \prod_{t=1}^{N}\left(1 - \eta_{src} \cdot \eta_{ch}^2 \cdot \eta_{QM}^2(t)\right) \quad (4)$$

The probability formulated in Eq.(4) corresponds to the probability of successfully distributing the entanglement between two nodes after $N$ timeslots, and therefore after having potentially sent $N$ pairs (we consider that the sources emit a pair of entangled photons by timeslot with a probability $\eta_{src}$). Generally speaking, when we look at a link budget, we look at the losses or the reception probabilities for a photon (or a pair of photons in this case). We must therefore divide the probability of Eq.(4) here by the number of modes in the memories $N$:

$$p_{success \geq 1(elem\ link)} = \frac{1 - \prod_{t=1}^{N}\left(1 - \eta_{src} \cdot \eta_{ch}^2 \cdot \eta_{QM}^2(t)\right)}{N} \quad (5)$$

A chain of entanglement switches is only a succession of elementary links, to which Bell state measurement modules are added (Figure 5-4).

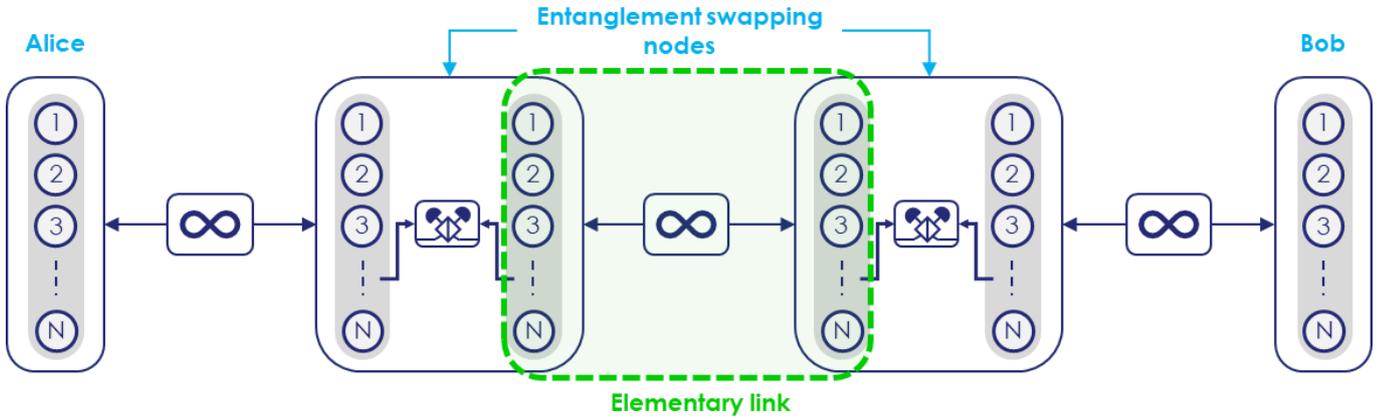

Figure 5-4: Quantum repeaters chain

Figure 5-4 shows a complete chain, with two end users Alice and Bob, as well as two quantum repeaters. A quantum repeater is defined as a Bell State Measurement module (BSM) and single photon detectors. For two repeater nodes we therefore have: 3 elementary links and 2 entanglement swapping nodes. Two new components appear in the performances of a repeaters chain:
- Bell State Measurement device (BSM);
- Single photon detectors.

The BSM is characterized by an efficiency $\eta_{BSM}$ related to the projection of the quantum state onto the good Bell state (this efficiency is necessarily less than 0.5). Single photon detectors are characterized by an efficiency $\eta_{det}$. A BSM requires two detectors simultaneously (four are required in total, but only two detectors are used per BSM during a Bell state measurement). By generalizing for $M$ entanglement switching nodes the chain in Figure 5-4, we count:
- $M$ quantum repeaters;

- $M$ +1 elementary links;
- $M$ Bell state measurements;
- $2M$ detectors for BSM measurements.

The probability of successfully transmitting at least one pair of entangled photons between the two end-users after $N$ timeslots is then:

$$p_{success \geq 1(chain)} = \frac{\left[1 - \prod_{t=1}^{N}\left(1 - \eta_{src} \cdot \eta_{ch}^2 \cdot \eta_{QM}^2(t)\right)\right]^{M+1} \cdot \eta_{BSM}^{M} \cdot \eta_{det}^{2M}}{N} \quad (6)$$

Since we consider identical entangled photon sources in the ground chain, with source rate $R_{src}$, the number of received pairs per second at the end-users is given by the formula:

$$N_{pair,\ ground} = R_{src} \cdot p_{success \geq 1(chain)} \quad (7)$$

Eq.(7) will be used for the calculation of the ground path performance in term of throughput.

The fidelity of the received pair at the user level is calculated with the following model, based on Werner states [26]. Similarly to the previous case on throughput modeling, we can model the evolution of fidelity by sections of elementary links. In order to evaluate the impact of unwanted photons on the overall fidelity of distributed states on the network, a decoherence channel model is assumed. For example, the states produced by the source are considered to be Werner states (noisy EPR pairs) [26] with an associated density matrix of the following form:

$$\rho_{src} = \mathcal{W}_{src} |\phi^+\rangle\langle\phi^+| + \frac{1 - \mathcal{W}_{src}}{4} \mathbb{I}_4 \quad (8)$$

where $|\phi^+\rangle$ is a perfectly entangled Bell state, $\mathcal{W}_{src}$ is the Werner parameter of the source associated with its fidelity $F_{src}$ with respect to $|\phi^+\rangle$, and $\mathbb{I}_4$ is a 4×4 identity matrix.

The Werner parameter of an elementary link can be estimated by concatenating the Werner parameters of each functional block. The source of entangled photons is characterized by $F_{src}$ fidelity because the latter may not produce perfect Bell states. We consider $\mathcal{W}_{src} = F_{src}$.

The noisy propagation channel is characterized by a fidelity $F_{fiber}$. Indeed, optical fiber induce polarization and phase rotations which are complicated to characterize and deviations in the measurement bases can be caused which will also impact the fidelity of the qubits. We consider $\mathcal{W}_{fiber} = F_{fiber}$.

Quantum memories are characterized by a $F_{QM}$ fidelity. When quantum memory re-emits a photon, its quantum state has interacted with the memory and is no longer exactly the same as when it was written. We consider $\mathcal{W}_{QM} = F_{QM}$.

We can model the Werner state of an elementary link $\mathcal{W}_{elem}$ as:

$$\mathcal{W}_{elem} = \mathcal{W}_{src} \cdot \mathcal{W}_{ch}^2 \cdot \mathcal{W}_{QM}^2 \quad (9)$$

The extension to this formula to a chain must take into account the Bell state measurement devices (BSM) including the single photon detectors. We will assume that only single photon detectors inside the BSM can have an impact on fidelity degradation, e.g. due to the dark counts. Single photon detectors are characterized by a "dark count" rate $R_{dc}$. The latter designates detections by a photodetector in the absence of incident light. They are caused by sources of noise intrinsic to the detector itself, such as thermal or electronic noise. A dark count detection can blind the detection of the photon of interest, which would induce a measurement error. To establish the effect of these dark counts on the fidelity, it is necessary to take into account the generation rate of the sources of entangled photons $R_{src}$ and the efficiency of the elementary links $p_{success\geq1(elem\ link)}$ as defined in Eq.(5), which will be written $p_{elem}$ in the following.

A click on a detector can come from two sources with different rates:
- click rate of detection of photon coming from the source $R_{true} = R_{src} \cdot p_{elem}$
- false click rate $R_{false} = R_{dc}$ (due to dark counts, etc. photon not coming from the source)

$R_{true}$ is the rate at which pairs of photons from the source, after attenuation in the channel and storage in memories, reach the detectors and are detected. This rate does not take into account pairs of photons for which one of the two photons would have been lost/absorbed.

The detector can either click because of a real photon or because of a dark count. The total click-through rate is therefore the sum of the two rates:

$$R_{total} = R_{true} + R_{false} = R_{src} \cdot p_{elem} + R_{dc} \qquad (10)$$

The probability of having a click of a photon coming from the source is given by:

$$P(true\ |click) = \frac{R_{true}}{R_{total}} = \frac{R_{src} \cdot p_{elem}}{R_{src} \cdot p_{elem} + R_{dc}} \qquad (11)$$

Eq.(11) can be used to obtain the Werner parameter for two connected elementary links labeled $i$ and $i+1$:

$$\mathcal{W}_{BSM,i} = \frac{R_{src} \cdot p_{elem,i}}{R_{src} \cdot p_{elem,i} + R_{dc}} \times \frac{R_{src} \cdot p_{elem,i+1}}{R_{src} \cdot p_{elem,i+1} + R_{dc}} \qquad (12)$$

Eq.(12) can be easily extrapolated to the terrestrial chain of nodes composed by $M$ elementary links and $M-1$ swapping nodes:

$$\mathcal{W}_{chain} = \prod_{i=1}^{M} \mathcal{W}_{elem,i} \cdot \prod_{i=1}^{M-1} \mathcal{W}_{BSM,i} \qquad (13)$$

Thus leading to:

$$\mathcal{W}_{chain} = \mathcal{W}_{src}^{M} \cdot \mathcal{W}_{ch}^{2M} \cdot \mathcal{W}_{QM}^{2M} \cdot \prod_{i=1}^{M-1} \left[ \frac{R_{src} \cdot p_{elem,i}}{R_{src} \cdot p_{elem,i} + R_{dc}} \cdot \frac{R_{src} \cdot p_{elem,i+1}}{R_{src} \cdot p_{elem,i+1} + R_{dc}} \right] \qquad (14)$$

The fidelity of the entire ground chain is thus [27]:

$$F_{ground} = \frac{1 + 3W_{chain}}{4} \tag{15}$$

**Model for the space path**:

The space path is composed of one satellite, two ground stations and two quantum memories (Figure 5-5). The number of entangled photon pairs received by two end-users is given by:

$$N_{pair,\ space} = R_{src} \cdot \eta_{src} \cdot \eta_A \cdot \eta_B \cdot \eta_{QM}^2(t_k), \tag{16}$$

with
- $R_{src}$ the entangle photon source rate
- $\eta_{src}$ the efficiency of the entangled photons source
- $\eta_A$ the channel efficiency of channel A
- $\eta_B$ the channel efficiency of channel B
- $\eta_{QM}(t_k)$ the efficiency of the quantum memory

The channel efficiencies are given by the space-to-ground link budget described hereafter. The received power of an optical signal sent from a satellite and received by a telescope at the ground is given by:

$$P_R = P_0 \cdot G_T \cdot \eta_{atm} \cdot G_R \tag{17}$$

with:

- $P_0$ the power of the entangled photon source
- $G_T = \eta_{cT}\left(1 - e^{-\frac{D_T^2}{2\omega_0^2}}\right)$ the transmitter gain (telescope efficiency). $\eta_{cT}$ is the internal transmittance of the on-board telescope (including wave front errors and obscuration), $D_T$ is the telescope aperture, and $\omega_0 = \frac{D_T}{2}$ is the beam waist at the output of the telescope.
- $\eta_{atm}$ the downlink atmospheric losses (absorption, scattering, cirrus clouds, aerosols). $\eta_{atm} = (\eta_{atm,0})^{\frac{1}{\sin(\theta)}}$ where $\theta$ is the satellite elevation and $\eta_{atm,0}$ the atmospheric losses when the satellite is at zenith ($\theta = \pi/2$).
- $G_R = \eta_{cR}\left(1 - e^{-\frac{D_R^2}{2\omega(r)^2}}\right)$ the receiver gain at the telescope level (telescope efficiency). $\eta_{cR}$ is the internal losses of the receiver telescope (including wave front errors, obscuration, optical fiber coupling with adaptive optics), $D_R$ is the telescope aperture, and $\omega(r)$ the received beam waist at the ground station side (with $\omega_0 = D_{sat}$), with $r$ the free space length between the satellite and the receiver's telescope pupil.

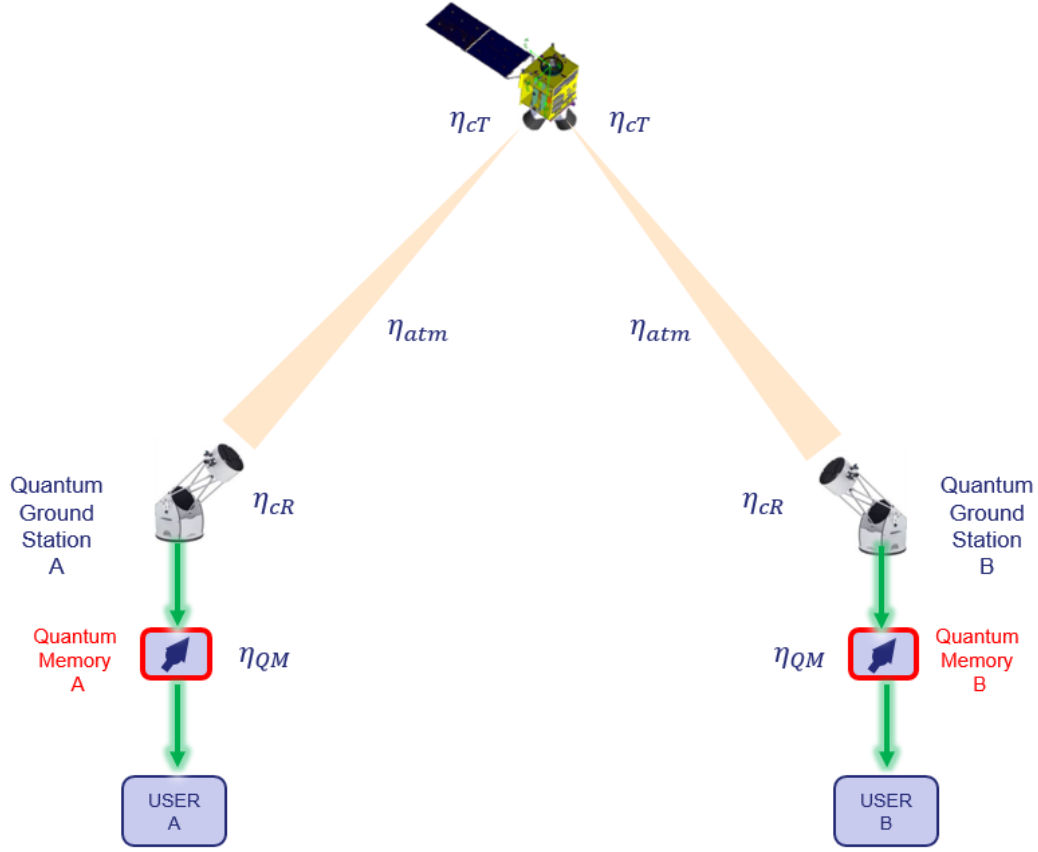

Figure 5-5: Space path connecting two users A and B, each user connected to a ground station with a quantum memory. The parameters $\eta$ represent the losses on each part of the channel (see text).

In gaussian optics, the beam waist is given by $\omega(r) = \omega_0\sqrt{1 + \left(\frac{r}{R_R}\right)^2}$, with the Rayleigh length $R_R = \frac{\pi \omega_0^2}{\lambda}$ ($\lambda$ is the signal wavelength).

Finally, the link budget is given by:

$$\eta_{A,B} = \frac{P_{R,A,B}}{P_0} = \eta_{cT} \cdot (1 - e^{-2}) \cdot \eta_{atm} \cdot \eta_{cR} \left(1 - e^{-\frac{2D_R^2}{D_T^2}\left(\frac{1}{1 + \frac{16\lambda^2 r_{A,B}^2}{\pi^2 D_T^4}}\right)}\right) \tag{18}$$

During a free-space link from a satellite, stray photons from the environment can be received in the telescopes at ground stations. These photons will have an impact on the fidelity of the quantum states, as they may be confused with useful photons sent by the satellite. The effect of these stray photons on the system's fidelity can be computed from the Coincidence-to-Accidental Ratio (CAR) between useful photons sent by the satellite and stray photons. The formula used is derived from [28] where they compute the Werner parameter of swapped states from a Bell State Measurement. In this model, we consider that the photons are directly stored in quantum memories and do not undergo entanglement swapping as the goal is only to distribute entanglement between the users receiving the photons. Ref. [28] gives a formula of the Coincidence-to-Accidental Ratio (CAR) as a function of stray light and expresses the Werner parameter as:

$$\mathcal{W}_{stray} \approx \frac{CAR - 1}{CAR + 1} \quad (19)$$

The space path is composed of a single elementary link with an entangled photons source, a propagation channel (freespace) and two quantum memories. Similarly to the ground path, the Werner parameter of this elementary link $\mathcal{W}_{elem}$ can be computed from Eq. (9).

The fidelity of the distributed photons over the space link can thus be computed using the Werner parameter of the elementary link and the Werner parameter of the contribution of stray photons:

$$\mathcal{W}_{space} = \mathcal{W}_{elem} \cdot \mathcal{W}_{stray} \quad (20)$$

We can then compute the fidelity of the space path, similarly to Eq. (15):

$$F_{space} = \frac{1 + 3\mathcal{W}_{space}}{4} \quad (21)$$

**List of parameters**:

Table 4 details the various parameters taken into account in our simulations.

**Table 4 : Parameters list user for simulations**

| Parameter | Symbol | Value |
|---|---|---|
| Wavelength | $\lambda$ | $1550\ nm$ |
| Bandwidth | $B_\lambda$ | $1\ nm$ |
| Source efficiency | $\eta_{src}$ | 0.1 |
| Source rate | $R_{src}$ | $1\ GHz$ |
| Source fidelity | $F_{src}$ | 0.99 |
| Quantum memory writting efficiency | $\eta_{QM}$ | 0.9 |
| Quantum memory fidelity | $F_{QM}$ | 0.99 |
| Quantum memory storage modes | $N_{QM}$ | 50 |
| Quantum memory characteristic storage time | $\tau_{QM}$ | $10\ ms$ |
| Quantum memory storage window | $\omega_{storage}$ | $250\ ps$ |
| Detector efficiency | $\eta_{det}$ | 0.9 |
| Detector dark count rate | $R_{dc}$ | $50\ cps$ |
| BSM efficiency | $\eta_{BSM}$ | 0.5 |
| Optical fiber attenuation | $\alpha$ | $0.2\ dB/km$ |
| Optical fiber link fidelity | $F_{fiber}$ | 0.99 |
| Tx telescope diameter | $D_{TX}$ | $30\ cm$ |
| Rx telescope diameter | $D_{RX}$ | $100\ cm$ |
| Rx telescope field of view | $FOV$ | $0.2\ \mu rad$ |
| Sky spectral radiance (broad-day value) | $I_{stray}$ | $0.3\ W/m^2/\mu m$ |

| Parameter | Symbol | Value |
|---|---|---|
| Free-space link fidelity | $F_{freespace}$ | 0.99 |
| Atmospheric losses when the satellite is at zenith | $\eta_{atm,0}$ | 0.5 |
| Internal transmittance of the ground telescope | $\eta_{cR}$ | 0.3 |
| Internal transmittance of the on-board telescope | $\eta_{cT}$ | 0.8 |
| Satellite altitude | | $600\ km$ |